\documentclass[a4, iop, numberedappendix]{emulateapj}

\usepackage{color}
\definecolor{red}{rgb}{1.00, 0.00, 0.00}
\definecolor{orange}{cmyk}{0.00, 0.40, 0.80, 0.20}
\definecolor{dark orange}{rgb}{0.71, 0.21, 0.01}
\definecolor{dark green}{rgb}{0.00, 0.39, 0.00}
\definecolor{dark blue}{rgb}{0.00,0.00,0.55}

\usepackage[colorlinks=true,linkcolor=blue,anchorcolor=blue,citecolor=blue,urlcolor=blue]{hyperref}

\usepackage{graphicx}
\usepackage{epsfig}
\usepackage{rotate}
\usepackage{afterpage}
\usepackage{amsmath}
\usepackage{amssymb}
\usepackage{multirow}
\usepackage{txfonts}
\usepackage{natbib}
\usepackage{blindtext}

\newcommand{\diff}{\mathrm{d}}
\newcommand{\scsc}[1]{{\scriptscriptstyle{#1}}}
\newcommand{\fref}[1]{{Figure.~\ref{#1}}}
\newcommand{\tref}[1]{{Table.~\ref{#1}}}

\def\aap{A\&A}			
\def\aj{AJ}				
\def\apj{ApJ}			
\def\apjl{ApJL}			
\def\apjs{ApJS}			
\def\jcap{JCAP}			
\def\mnras{MNRAS}		
\def\nat{Nature}			
\def\physrep{Phys.~Rep.}	
\def\prd{Phys.~Rev.~D}	

\begin{document}

\title{Void profile from \textit{Planck} lensing potential map}
\author{Teeraparb Chantavat\altaffilmark{1 $\dagger$}}\thanks{$\dagger$ E-mail: teeraparbc@nu.ac.th}
\author{Utane Sawangwit\altaffilmark{2}}
\author{Benjamin D. Wandelt\altaffilmark{3, 4, 5}}

\altaffiltext{1}{Laboratory of Cosmology and Gravity, The Institute for Fundamental Study, Naresuan University ``The Tah Poe Academia Institute", Naresuan University, Phitsanulok 65000, Thailand}
\altaffiltext{2}{National Astronomical Research Institute of Thailand (NARIT), Chiang Mai, 50200, Thailand}
\altaffiltext{3}{Sorbonne Universit\'{e}, UPMC Univ Paris 6 et CNRS, UMR 7095, Institut d'Astrophysique de Paris, 98 bis bd Arago, 75014 Paris, France}
\altaffiltext{4}{Sorbonne Universit\'{e}, Institut Lagrange de Paris (ILP), 98 bis Boulevard Arago, 75014 Paris, France}
\altaffiltext{5}{Departments of Physics and Astronomy, University of Illinois at Urbana-Champaign, Urbana, IL 61801, USA}

\date{\today}

\begin{abstract}

We use the lensing potential map from \textit{Planck} CMB lensing reconstruction analysis and the ``Public Cosmic Void Catalog" to measure the stacked void lensing potential.  In this profile, four parameters are needed to describe the shape of voids with different characteristic radii $R_V$.  However, we have found that after reducing the background noise by subtracting the average background, there is a residue lensing power left in the data.  The inclusion of the environment shifting parameter, $\gamma_V$, is necessary to get a better fit to the data with the residue lensing power.  We divide the voids into two redshift bins: cmass1 ($0.45 < z < 0.5$) and cmass2 ($0.5 < z < 0.6$).  Our best-fit parameters are $\alpha = 1.989\pm0.149$, $\beta = 12.61\pm0.56$, $\delta_c=-0.697\pm0.025$,  $R_S/R_V=1.039\pm0.030$, $\gamma_v=(-7.034\pm0.150) \times 10^{-2}$ for the cmass1 sample with 123 voids and $\alpha = 1.956\pm0.165$, $\beta = 12.91\pm0.60$, $\delta_c=-0.673\pm0.027$,  $R_S/R_V=1.115\pm0.032$, $\gamma_v=(-4.512\pm0.114) \times 10^{-2}$ for the cmass2 sample with 393 voids at 68\% C.L.  The addition of the environment parameter is consistent with the conjecture that the Sloan Digital Sky Survey voids reside in an underdense region.
\end{abstract}


\section{Introduction}
\label{sec:introduction}

In the standard cosmological model, the universe is homogeneous and isotropic on large scales.  The seeds of present-day large-scale structure of the universe are formed from the highly Gaussian and nearly scale-invariant power spectrum of matter density \citep{Hinshaw_ea2013, PlanckXIII_2015}.  However, on small scales, the hierarchical clustering of matter leads to formations of complex cosmic structure such as clusters of galaxies, walls, filaments, and voids \citep{Boylan-Kolchin_ea2009}.  Among all of the large-scale objects in the universe,  cosmic voids, which are large underdensities in the matter distribution, occupy the vast majority of the universe and hence provide the largest volume-based test on theories of structure formation \citep{Ceccarelli_ea2006}.  Being interesting objects in their own right, they contain a wealth of information on the fundamental properties of the universe.  For example, the low-density environment of voids is a perfect place to study galaxies, as the galaxies are expected not to be affected by the complex astrophysical processes that modify galaxies in high-density environments and allows galaxies to evolve independently without environmental effects \citep{Beygu_ea2013, Penny_ea2015}.  In addition, since voids occupy the cosmic volume where the matter density is lowest, the difference between dark energy and modified gravity models for cosmic acceleration could be distinguishable within cosmic voids \citep{Clampitt_ea2013, Cai_ea2015, Barreira_ea2015, Zivick_ea2015}.  

The computational approach called $N$-body dark matter simulations is one of the best tools to empirically understand various void properties such as number functions \citep{Sheth_vandeWeygaert2004, Jennings_ea2013} and void ellipticity functions \citep{Biswas_ea2010}.  However, the definition of voids is rather vague, and various definitions exist in the literature; some are more suitable for theoretical calculations, while the others are more suitable for observations or $N$-body simulations.  This variety of definitions renders comparison of theoretical predictions on void properties and observations difficult.  \texttt{ZOBOV} (Zones Bordering On Voidness; \cite{Neyrinck_2008}) and \texttt{WVF} (Watershed Void Finder;\cite{Platen_ea2007}) are two of the popular void-finding algorithms.  Both methods are based on some tessellation methods and the watershed concept of defining voids.  \texttt{ZOBOV} requires no free parameters or assumptions about the shape and is based on Voronoi tessellation.  However, the \texttt{ZOBOV} voids are unsmooth and rather edgy.  \texttt{WVF} also requires no free parameters and is based on a watershed transform.  However, \texttt{WVF} uses several techniques to smooth the density field so that the \texttt{WVF} voids are not edgy.  With void identification algorithms being progressively developed for galaxy redshift surveys such as the Sloan Digital Sky Survey (SDSS), cosmic voids are being continually found, amounting to releases of public voids catalogs \citep{Pan_ea2012, Sutter_ea2012a, Sutter_ea2014a, Nadathur_2016}.

Recently, there has be an increasing amount of attention on voids as objects for various aspects of cosmological studies.  The dynamic of voids and redshift-space distortion \citep{Kaiser_1987} is one of the probes of the growth of large-scale structure.  The amount of dark matter in the universe could be obtained from the peculiar velocity fields \citep{Courtois_ea2012}.  The relationship between their extent angular size and the distance along the line of sight, known as the Alcock--Paczy\'{n}ski test \citep{Alcock_Paczynski1979}, is predicted to be a promising probe of dark energy by using a stacking method to obtain a statistically averaged shape of voids in 2D or 3D spaces.  By stacking a large number of voids, one would expect the difference in radial and transverse direction to be directly related to the product of angular distance and the Hubble parameter \citep{Lavaux_Wandelt2012, Sutter_ea2012b, Sutter_ea2014b}.  Another geometrical study, the evolution of the ellipticity of voids, could be used as a tool in practice to constrain the dark energy equation of state \citep{Lee_Park2009, Bos_ea2012}.  The Integrated Sachs--Wolfe (ISW) effect \citep{Sachs_Wolfe1967} caused by the evolution of the gravitational potential within voids can also be detected \citep{Granett_ea2008, Cai_ea2014, Chen_Kantowski2015, Hotchkiss_ea2015, Ilic_ea2013, PlanckXXI_2015}.

The measurement of weak gravitational lensing probes the matter distribution by means of the deflection of light from the background sources.  The trajectories of photons from background sources are bent toward gravitating matter due to the distortion of spacetime caused by gravitational lensing \citep{Einstein_1936}.  The scenario is reversed when voids are acting as the sources of gravitational lenses instead of dark matter.  The delensing effect of voids has been investigated and recently observed through the distortions of background galaxies by a stacking method that enhances the signal \citep{Higuchi_ea2013, Krause_ea2013, Melchior_ea2014, Clampitt_Jain2015, Gruen_ea2016}.

\setcounter{footnote}{0}

The cosmic microwave background (CMB) radiation, which is the signal from surface of the last scattering surface, exists as the ubiquitous background for gravitational lensing.  The gravitational anti-lensing effect of voids has been recently investigated by \citep{Bolejko_ea2013}, \citep{Chen_ea2015}, \citep{Das_Spergel2009}. The CMB signals that are lensed by multiple voids are also a promising tool to obtain good constraints on cosmological parameters \citep{Chantavat_ea2016}. \textit{Planck} \citep{PlanckXV_2015} has released lensing potential maps from the CMB\footnote{Can be downloaded from \href{http://www.sharelatex.com}{http://pla.esac.esa.int/}} that utilized quadratic estimators that exploit the statistical anisotropy induced by lensing \citep{Okamoto_Hu2003}.  From a theoretical point of view, if the matter density distribution within a void, known as a void profile, is known, then the lensing potential could be computed and vice versa.  The statistical average void density profile known as the universal void profile (here after, HSW; \cite{Hamaus_ea2014a}) has been released and potentially could be exploited to predict the lensing effect of voids at various sizes and redshifts with only a few parameters (more in \S\ref{sec:theory}).  Hence, from the lensing potential data from \textit{Planck}, one could, in principle, derive the HSW parameters.  This would be a good consistency check if the HSW void profile could be reverse-engineered from observables.

The goals of this article are (1) to extract the stacked lensing potential from \textit{Planck} lensing data by cross-correlation with voids from SDSS data \citep{Sutter_ea2014b}, (2) to compare and cross-examine the derived HSW void parameters from the \textit{Planck} lensing potential map with other methods, and (3) to better understand the effect of gravitational lensing from voids in the \textit{Planck} lensing data.  We shall begin with a description of the void catalog from the SDSS galaxy redshift survey \citep{Sutter_ea2014b} and \textit{Planck} lensing data \citep{PlanckXV_2015} in \S\ref{sec:data}.  The extraction and cross-correlation methods are also discussed.  In \S\ref{sec:theory}, we describe the HSW void profile parameterization, and a brief overview of the gravitational lensing effect with voids is also introduced.  Our parameter estimation is described in \S\ref{sec:paraest}, and the results are shown in \S\ref{sec:results}.  The discussions and conclusions are given in \S\ref{sec:discussion}.  Throughout this article, our fiducial cosmological parameters are $\Omega_M = 0.315$, $\Omega_\Lambda = 0.685$, $H_0 = 67.3\ \text{km s}^{-1} \text{Mpc}^{-1}$, $w = -1$, and $\Omega_k = 0$, which is consistent with a flat $\Lambda$CDM cosmology from \textit{Planck} 2013 + WMAP polarization maximum likelihood cosmological parameters \citep{PlanckXVI}.

\section{Data and Methodology}
\label{sec:data}
This study utilizes the lensing potential measured from the lensed CMB maps to extract the potential in the vicinity around cosmic voids.  Here, we describe the datasets and method for measuring the stacked voids' lensing potential, which will be used to constrain the void density profile, as discussed further in \S\ref{sec:theory}.

\subsection{\textit{Planck} Lensing Potential map}
\label{ssec:planckdata}
The descriptions of data products and an overview of the scientific results of the \textit{Planck} full-mission data release are given by \citet{PlanckI_2015}.  Apart from CMB temperature, polarization frequency maps, and foreground component maps, the science team also released the 
reconstructed lensing potential map of the CMB \citep{PlanckXV_2015}.  They applied quadratic lensing estimators and procedures described by \citet{Okamoto_Hu2003} to the foreground-cleaned CMB map.  The foreground-cleaned map was constructed from all the frequency band maps using the \texttt{SMICA} procedure \citep{PlanckIX_2015}.  The contaminated regions of the \texttt{SMICA} map were further removed with the Galaxy, point-source, and \texttt{SMICA}-specific temperature and polarization masks, which leave $\approx\!67\%$ of the sky for the minimum-variance lensing reconstruction analysis.  Given the sensitivity and angular resolutions of the \textit{Planck} mission, the analysis thus results in the most significant measurement of the CMB lensing potential map to date. 

\begin{figure}
	\hspace{-2mm}
	\includegraphics[scale=0.35,angle=90]{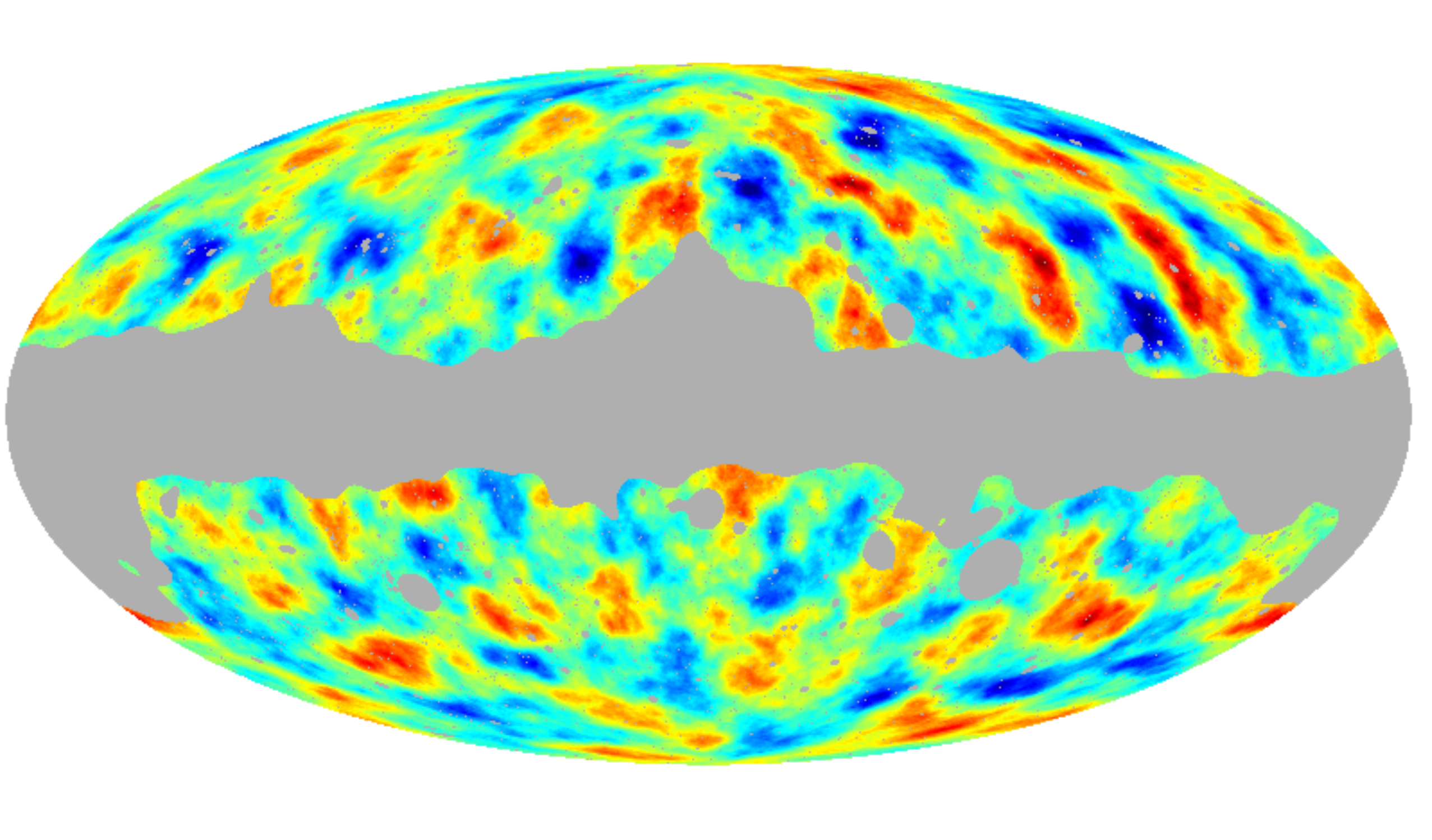}
	\caption{\label{fig:potentialmap}  Lensing potential map, $\psi(\hat{n})$, constructed from \textit{Planck} lensing convergence, $\kappa_{\ell m}$.  The gray shaded area marks the rejected pixels excluded by the analysis mask.}
\end{figure}

The online data provided by the \textit{Planck} science team are in the standard \texttt{HEALPix} format \citep{HEALPixref}.  The spherical harmonics coefficients of lensing convergence, $\kappa_{\ell m}$, are given for multipoles up to $\ell_{\rm max} = 2048$ instead of the lensing potential, $\psi_{\ell m}$. We use the standard definition of lensing convergence to calculate  $\psi_{\ell m}$ from 
\begin{equation}
	\label{eq:convergence}
	\kappa_{\ell m} =\frac{\ell (\ell +1)}{2} \psi_{\ell m}.
\end{equation}
We then use the \texttt{HEALPix} \texttt{synfast} package to synthesize the lensing potential map, $\psi(\hat{n})$.  The generated map has a resolution of $N_{\rm side} = 2048$ (i.e. $\approx 1'.7 \times 1'.7$ pixels). The projected lensing potential map is shown in Figure.~\ref{fig:potentialmap}.  This reconstructed lensing potential map is used to cross-correlate with the cosmic voids catalog, where the azimuthally averaged potential around each void is extracted, scaled, and then combined to constrain the void density profile.  For our analysis, we shall use only those regions that pass the analysis mask (Galaxy + point-source + \texttt{SMICA}). 

\subsection{Cosmic voids catalogue}
\label{ssec:voidcatalog}
In this work, we use the ``Public Cosmic Void Catalog" \citep{Sutter_ea2012a} constructed using a modified and extended version of the watershed algorithm \texttt{ZOBOV}, called `Void IDentification and Examination' \citep[VIDE;][]{Sutter_ea2014b}.  We applied the code to the SDSS Data Release 7 \citep{SDSSDR7} main galaxy sample and SDSS-III Baryon Oscillation Spectroscopic Survey (BOSS) Data Release 10 \citep{SDSSDR10} LOWZ and CMASS samples.  This results in $\approx1\,500$ individually detected voids.  The samples represent volume-limited catalogs of voids at various redshift bins.

To minimize any possible evolution of the  profile parameters, we choose to work with voids in individual redshift bins and do not stack voids across the bins.  In order to have as many voids in a single bin as possible, we chose a high-redshift bin to increase the volume.  Here, we report on the analysis using the ``dr10cmass1" ($0.45 < z < 0.5$, $\Delta t_{\rm age}=0.4 {\rm ~Gyrs}$, $N_{\rm void}=229$) and the ``dr10cmass2" samples ($0.5 < z < 0.6$, $\Delta t_{\rm age}=0.7 {\rm ~Gyrs}$, $N_{\rm void}=696$), hereafter called ``cmass1" and ``cmass2," respectively.

\begin{figure}
	\hspace{-0.7cm}
	\includegraphics[scale=0.45]{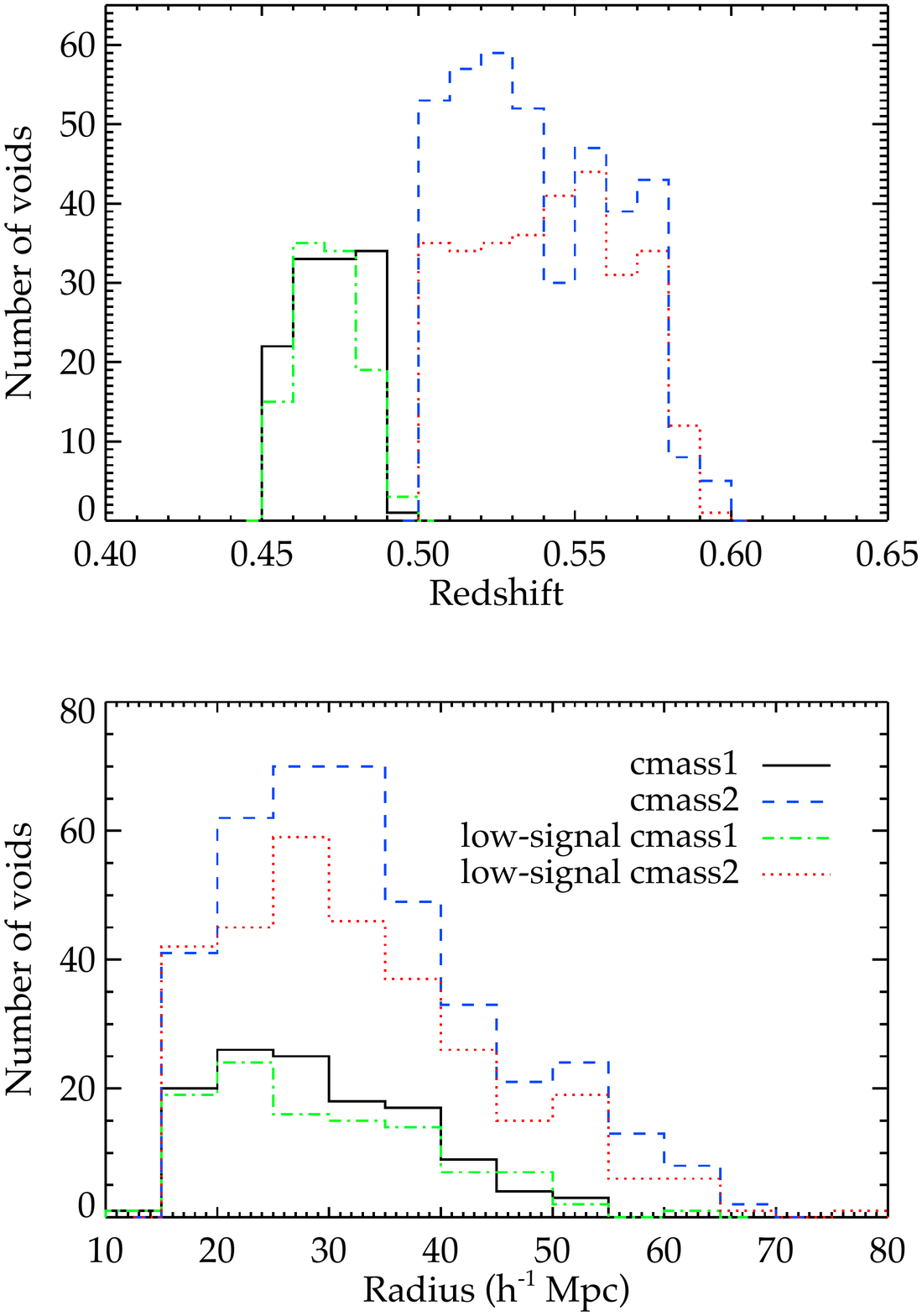}
	\caption{\label{fig:redshiftdist}Redshift (top) and radius (bottom) distributions of the void samples used in our analyses from the ``dr10cmass1" (black solid lines) and ``dr10cmass2" (blue dashed lines) samples, where their medians are $z_{\rm{cmass1}} = 0.47$, $R_{V,\rm{cmass1}}=28.26 ~h^{-1}$Mpc and $z_{\rm{cmass2}} = 0.53$, $R_{V,\rm{cmass2}}=31.68 ~h^{-1}$Mpc, respectively.  For comparison, the redshift and radius distributions of voids rejected for the stacking analysis, as a result of a very weak lensing signal or having negative lensing potential, are also shown.  We do not observe any significant deviation in their redshift and radius distributions.}
\end{figure}

We therefore stacked the \textit{Planck} lensing potential map (\S \ref{ssec:planckdata}) around the central positions of cosmic voids taken from the \cite{Sutter_ea2014b} ``dr10cmass1" and  ``dr10cmass2" samples as two separated measurements.  However, during our analysis we found that considerable numbers (106 and 303 for cmass1 and cmass2, respectively) of the voids are located in the negative lensing potential regions, which could be due to a number of reasons and warrants further investigations.  We therefore restrict our analysis to 123 and 393 voids from samples cmass1 and cmass2, respectively.  The radius and redshift distributions of the subsamples that we used do not show any significant difference from those of the excluded low-signal subsamples (see Figure.~\ref{fig:redshiftdist}).

\subsection{Stacking analysis}
\label{ssec:stack}
The stacking analysis is performed on the \textit{Planck} lensing potential map (See \S\ref{ssec:planckdata}) around each of the voids' centre.  Since we need to compare our measurements to the theoretical prediction of the size-independent potential, $\tilde{\psi}(r/R_V)$,  (see Eq.~(\ref{eq:scale_potential}) in \S \ref{ssec:voidlensing}), but the potential map is a 2D projection on the surface of a sphere, we therefore bin up the lensing potential according to physical separation and not angular separation.  Therefore, for each void, we use the comoving angular diameter distance, $D_{A}(z)$, to scale the pixels of the lensing maps surrounding the vicinity of each void according to its redshift $z$ and radius $R_{V}$. For the $i$th $(r/R_V)_i$ bin, its corresponding angular bin is given by
\begin{equation}
	\theta_i \left(z,R_V\right) = R_V / D_{A}\left(z\right) \times (r/R_V)_i.
\end{equation}
The lensing potential value in each $r/R_{V}$ bin around a void is azimuthally averaged and is then background subtracted by the measurement at the largest separation, $10\,R_V$.  Next, we scale the amplitude of the overall lensing potential measured from each void according to Eq.~(\ref{eq:scalefactor}) to account for their different redshifts and radii before we combine them together.  Each lensing potential is scaled to the its respective median redshift and radius,
\begin{equation}
	\label{eq:lensingData}
	\tilde\psi\left(r/R_V\right)_i= \psi\left(r; z,R_V\right)_i  \times \frac{R_{V,\rm{med}}^3 (1+z_{\rm med})^3 D_{+}(z_{\rm med})D_{A}(z)}{R_{V}^3 (1+z)^3D_{+}(z)D_{A}(z_{\rm med})},
\end{equation}
where $R_{V,\rm{med}}$ and $z_{\rm med}$ are the median void radius and redshift, respectively, of the bins given in Figure.~\ref{fig:redshiftdist}.  The average lensing potential is then calculated from the scaled measurements of 123 and 393 voids for cmass1 and cmass2 samples, respectively.  The uncertainties of our measurement are then calculated using the \textit{jackknife} resampling technique.  The void sample is separated into 12 subsamples according to their Galactic latitudes and longitudes.  We then remeasure the lensing potential 12 times, each $j$th time leaving out one subsample, and the \textit{jackknife} error is given by
\begin{eqnarray}
	\label{eq:jackknife}
	\nonumber \Sigma^{\rm JK}_{ij}\left(r/R_V\right) &=& \frac{11}{12}\sum^{12}_{i, j}\Big(\tilde\psi_i\left(r/R_V\right)-\bar{\psi}\left(r/R_V\right)\Big) \\
	&& \times \Big(\tilde\psi_j\left(r/R_V\right)-\bar{\psi}\left(r/R_V\right)\Big),
\end{eqnarray} 
where $\bar{\psi}\left(r/R_V\right)$ is the averaged lensing potential from 12 \textit{jackknife} subsamples.  The measured lensing potential and the estimated uncertainties are shown in Figure.~\ref{fig:lensp_long}.  Using the same \textit{jackknife} sub-sampling method, we also estimate the correlations between measurements of different bins (off-diagonal elements). The estimated covariance matrices are then used in our fitting procedure as described in \S\ref{sec:paraest}.

\section{Theory}
\label{sec:theory}

In this section, we shall describe all the relevant theories in this analysis, such as the parameterization of the void density profile and the theory of gravitational lensing potential applicable to voids.

\subsection{Void Density Profile Parameterization}
\label{ssec:voidprofile}

In general, voids will be observed with various shapes and orientations in the field of view.  However, the averaged void density profile will be spherically symmetric and is well fitted by the universal void density profile \citep{Hamaus_ea2014a}.  The profile is given by
\begin{equation}
	\frac{\delta\rho_V(r)}{\bar{\rho}_M} = \delta_c \frac{1 - \left(r/R_S\right)^\alpha}{1 + (r/R_V)^\beta} + \gamma_V,
\end{equation}
where $\bar{\rho}_M$ is the mean cosmic matter density and $\delta\rho_V$ is the density deviation for the mean density.  $R_V$ is the characteristic void radius, and $R_S$ is a scale radius where $\rho_V(R_S) = \bar{\rho}_M$.  $\alpha, \beta$ and $\delta_c$ are the shape parameters.  $\gamma_V$ is an additional environment shifting parameter that was not included in the original model.  However, the benefit of the inclusion of the additional parameter is twofold.  First, the parameter takes into account of any systematic uncertainties that may occur in the data extraction process.  The parameter $\gamma_V$ is considered as a nuisance parameter, which will be marginalized later.  Second, there is a tendency that voids in the SDSS catalogs may reside in an underdense region of the universe \citep{Hamaus_ea2014b}.  The constant shifting parameter will take the locality of the environment of voids into account.  We shall take the average void profile as our estimate of the void profile in the analysis.  Since $R_V$ has been given from the data, we shall use $R_S/R_V$ as one of the fitting parameters.  Hence, our vector in the parameter space is $\boldsymbol{X} = \left\{ \alpha, \beta, \delta_c, R_S/R_V, \gamma_V \right\}$.

\subsection{Void Gravitational Lensing Potential}
\label{ssec:voidlensing}

We encourage readers to consult \cite{Bartelmann_Schneider2001} for a general review of gravitational weak lensing.  The gravitational potential at redshift $z$ is given by
\begin{equation}
	\label{eq:poisson}
	\nabla^2 \Psi_{\scsc{N}} = 4 \pi G \bar{\rho}_{M0} (1 + z) D_{+}(z)\delta_M(z = 0),
\end{equation}
where $D_{+}(z)$ is the growth function normalized to unity at $z = 0$.  $\rho_{M0}$ is the matter density at the present epoch.  The lensing potential is defined as the integral over the line-of-sight direction $\hat{n}$,
\begin{eqnarray}
	\label{eq:voidlensing}
	\psi(\hat{n}) = -\frac{2}{c^2} \int \mbox{d}\chi\ \boldsymbol{\nabla}_\perp \Psi_{\scsc{N}}(\chi\hat{n}),
\end{eqnarray}
where $\chi$ is the comoving distance.  $\boldsymbol{\nabla}_\perp$ is the transverse derivative
\begin{equation}
	\boldsymbol{\nabla}_\perp \equiv \boldsymbol{\hat{\theta}}\frac{\partial}{\partial\theta} + \frac{\boldsymbol{\hat{\phi}}}{\sin\theta}\frac{\partial}{\partial\phi}\ .
\end{equation}
  The excess surface density is given by the line-of-sight integral
\begin{equation}
	\delta\tilde{\sigma}_{V}(b) = \int_{-\infty}^{\infty}\diff x\ \delta\rho_V\left(\sqrt{x^2 + b^2}\right),
\end{equation}
where $x = r/R_V$ and $b$ is the scaled impact parameter.  It is convenient to define scale-invariant quantities that are a function of the scaled radius $x$.  In this article, all the scale-invariant quantities are denoted by the tilde symbol.

From Eq.~(\ref{eq:voidlensing}), the lensing potential is given by
\begin{equation}
	\tilde{\psi}(b) = \int_0^b \frac{\diff y}{y} \int_0^{y}\diff x\ x \ \delta\tilde{\sigma}(x).
\end{equation}
The lensing potential will be given by
\begin{eqnarray}
\label{eq:scale_potential}
	\psi(r; R_V, z) = \mathcal{S}(R_V, z)\times\tilde{\psi}(r/R_V),
\end{eqnarray}
where the scaling factor $\mathcal{S}(R_V, z)$ is
\begin{eqnarray}
	\nonumber\mathcal{S}(R_V, z) &=& \frac{16\pi G}{c^2} \bar{\rho}_{m0} \left(\frac{R_V}{\mbox{Mpc}\ h^{-1}}\right)^3 \times \frac{(1 + z)^3 D_{+}(z)}{\left(D_A(z) / \mbox{Mpc}\ h^{-1}\right)},\\
	\label{eq:scalefactor}
\end{eqnarray}
where $D_A(z)$ is the comoving angular diameter distance.

\begin{table*}\centering
\caption{\label{tab:fittingparams} Fitting parameters for Model A and Model B.}
\begin{tabular}{|p{0.18\columnwidth}|p{0.095\columnwidth}p{0.03\columnwidth}p{0.01\columnwidth}|p{0.095\columnwidth}p{0.03\columnwidth}p{0.01\columnwidth}|p{0.095\columnwidth}p{0.03\columnwidth}p{0.01\columnwidth}|p{0.095\columnwidth}p{0.03\columnwidth}p{0.01\columnwidth}|p{0.095\columnwidth}p{0.03\columnwidth}p{0.01\columnwidth}|p{0.095\columnwidth}p{0.03\columnwidth}p{0.01\columnwidth}|}
	\hline
	\multicolumn{19}{|c|}{CMASS 1 SAMPLE} \\
	\hline
	\hfil \multirow{3}{*}{Parameters} & \multicolumn{6}{|c|}{Data-I}&\multicolumn{6}{|c|}{Data-II} & \multicolumn{6}{|c|}{Data-III} \\
	\cline{2-19}
	& \multicolumn{3}{|c|}{Model A} & \multicolumn{3}{|c|}{Model B} & \multicolumn{3}{|c|}{Model A} & \multicolumn{3}{|c|}{Model B} & \multicolumn{3}{|c|}{Model A} & \multicolumn{3}{|c|}{Model B} \\
	& \multicolumn{3}{|c|}{(without $\gamma_V$)} & \multicolumn{3}{|c|}{(with $\gamma_V$)} & \multicolumn{3}{|c|}{(without $\gamma_V$)} & \multicolumn{3}{|c|}{(with $\gamma_V$)} & \multicolumn{3}{|c|}{(without $\gamma_V$)} & \multicolumn{3}{|c|}{(with $\gamma_V$)} \\
	\hline
	\hfil $\alpha$ & \hfill $2.471$ & \hspace{-1em}$\pm$ &\hspace{-2.4em}$0.127$ & 
	\hfill $2.016$ & \hspace{-1em}$\pm$ & \hspace{-2.4em}$0.105$ &
	\hfill $2.612$ & \hspace{-1em}$\pm$ & \hspace{-2.4em}$0.116$ &
	\hfill $1.923$ & \hspace{-1em}$\pm$ & \hspace{-2.4em}$0.130$ &
	\hfill $2.642$ & \hspace{-1em}$\pm$ & \hspace{-2.4em}$0.107$ &
	\hfill $1.989$ & \hspace{-1em}$\pm$ & \hspace{-2.4em}$0.149$ \\
	\hfil $\beta$ & \hfill $12.35$&\hspace{-1em}$\pm$&\hspace{-2.4em}$0.62$ &
	\hfill $11.59$ & \hspace{-1em}$\pm$ & \hspace{-2.4em}$0.45$ & 
	\hfill $13.53$ & \hspace{-1em}$\pm$ & \hspace{-2.4em}$0.49$ &
	\hfill $11.68$ & \hspace{-1em}$\pm$ & \hspace{-2.4em}$0.48$ &
	\hfill $13.92$ &\hspace{-1em}$\pm$&\hspace{-2.4em}$0.51$ &
	\hfill $12.61$ & \hspace{-1em}$\pm$ & \hspace{-2.4em}$0.56$ \\
	\hfil $\delta_c$ & \hfill $-0.739$&\hspace{-1em}$\pm$&\hspace{-2.4em}$0.024$ 	& 
	\hfill $-0.699$ & \hspace{-1em}$\pm$ & \hspace{-2.4em}$0.018$ &
	\hfill $-0.766$&\hspace{-1em}$\pm$&\hspace{-2.4em}$0.022$ &
	\hfill $-0.689$ & \hspace{-1em}$\pm$ & \hspace{-2.4em}$0.024$ &
	\hfill $-0.786$&\hspace{-1em}$\pm$&\hspace{-2.4em}$0.020$ &
	\hfill $-0.697$ & \hspace{-1em}$\pm$ & \hspace{-2.4em}$0.025$ \\
	\hfil $R_S/R_V$ & \hfill $1.148$&\hspace{-1em}$\pm$&\hspace{-2.4em}$0.025$ & 
	\hfill $0.969$ & \hspace{-1em}$\pm$ & \hspace{-2.4em}$0.028$ &
	\hfill $1.199$&\hspace{-1em}$\pm$&\hspace{-2.4em}$0.021$ &
	\hfill $1.006$ & \hspace{-1em}$\pm$ & \hspace{-2.4em}$0.034$ &
	\hfill $1.236$&\hspace{-1em}$\pm$&\hspace{-2.4em}$0.021$&
	\hfill $1.039$ & \hspace{-1em}$\pm$ & \hspace{-2.4em}$0.030$ \\
	\hfil $\gamma_V\ (\times\ 10^{-2})$	& \multicolumn{3}{|c|}{N/A} &
	\hfill $-9.679$&\hspace{-1em}$\pm$&\hspace{-2.4em}$0.663$ &
	\multicolumn{3}{|c|}{N/A} &
	\hfill $-8.505$&\hspace{-1em}$\pm$&\hspace{-2.4em}$0.298$ &
	\multicolumn{3}{|c|}{N/A} &
	\hfill $-7.034$&\hspace{-1em}$\pm$&\hspace{-2.4em}$0.150$ \\
	\hline
	\multicolumn{19}{|c|}{CMASS 2 SAMPLE} \\
	\hline
	\hfil \multirow{3}{*}{Parameters} & \multicolumn{6}{|c|}{Data-I}&\multicolumn{6}{|c|}{Data-II} & \multicolumn{6}{|c|}{Data-III} \\
	\cline{2-19}
	& \multicolumn{3}{|c|}{Model A} & \multicolumn{3}{|c|}{Model B} & \multicolumn{3}{|c|}{Model A} & \multicolumn{3}{|c|}{Model B} & \multicolumn{3}{|c|}{Model A} & \multicolumn{3}{|c|}{Model B} \\
	& \multicolumn{3}{|c|}{(without $\gamma_V$)} & \multicolumn{3}{|c|}{(with $\gamma_V$)} & \multicolumn{3}{|c|}{(without $\gamma_V$)} & \multicolumn{3}{|c|}{(with $\gamma_V$)} & \multicolumn{3}{|c|}{(without $\gamma_V$)} & \multicolumn{3}{|c|}{(with $\gamma_V$)} \\
	\hline
	\hfil $\alpha$ & \hfill $2.580$ & \hspace{-1em}$\pm$ &\hspace{-2.4em}$0.129$ & 
	\hfill $1.935$ & \hspace{-1em}$\pm$ & \hspace{-2.4em}$0.119$ &
	\hfill $2.687$ & \hspace{-1em}$\pm$ & \hspace{-2.4em}$0.116$ &
	\hfill $1.617$ & \hspace{-1em}$\pm$ & \hspace{-2.4em}$0.126$ &
	\hfill $2.959$ & \hspace{-1em}$\pm$ & \hspace{-2.4em}$0.122$ &
	\hfill $1.956$ & \hspace{-1em}$\pm$ & \hspace{-2.4em}$0.165$ \\
	\hfil $\beta$ & \hfill $12.22$&\hspace{-1em}$\pm$&\hspace{-2.4em}$0.52$ &
	\hfill $11.43$ & \hspace{-1em}$\pm$ & \hspace{-2.4em}$0.49$ & 
	\hfill $13.85$ & \hspace{-1em}$\pm$ & \hspace{-2.4em}$0.50$ &
	\hfill $12.32$ & \hspace{-1em}$\pm$ & \hspace{-2.4em}$0.52$ &
	\hfill $13.71$ &\hspace{-1em}$\pm$&\hspace{-2.4em}$0.53$ &
	\hfill $12.91$ & \hspace{-1em}$\pm$ & \hspace{-2.4em}$0.60$ \\
	\hfil $\delta_c$ & \hfill $-0.726$&\hspace{-1em}$\pm$&\hspace{-2.4em}$0.021$ 	& 
	\hfill $-0.668$ & \hspace{-1em}$\pm$ & \hspace{-2.4em}$0.019$ &
	\hfill $-0.784$&\hspace{-1em}$\pm$&\hspace{-2.4em}$0.022$ &
	\hfill $-0.657$ & \hspace{-1em}$\pm$ & \hspace{-2.4em}$0.021$ &
	\hfill $-0.817$&\hspace{-1em}$\pm$&\hspace{-2.4em}$0.020$ &
	\hfill $-0.673$ & \hspace{-1em}$\pm$ & \hspace{-2.4em}$0.027$ \\
	\hfil $R_S/R_V$ & \hfill $1.172$&\hspace{-1em}$\pm$&\hspace{-2.4em}$0.023$ & 
	\hfill $0.991$ & \hspace{-1em}$\pm$ & \hspace{-2.4em}$0.031$ &
	\hfill $1.260$&\hspace{-1em}$\pm$&\hspace{-2.4em}$0.017$ &
	\hfill $1.002$ & \hspace{-1em}$\pm$ & \hspace{-2.4em}$0.037$ &
	\hfill $1.276$&\hspace{-1em}$\pm$&\hspace{-2.4em}$0.016$&
	\hfill $1.115$ & \hspace{-1em}$\pm$ & \hspace{-2.4em}$0.032$ \\
	\hfil $\gamma_V\ (\times\ 10^{-2})$	& \multicolumn{3}{|c|}{N/A} &
	\hfill $-5.716$&\hspace{-1em}$\pm$&\hspace{-2.4em}$0.693$ &
	\multicolumn{3}{|c|}{N/A} &
	\hfill $-6.747$&\hspace{-1em}$\pm$&\hspace{-2.4em}$0.239$ &
	\multicolumn{3}{|c|}{N/A} &
	\hfill $-4.512$&\hspace{-1em}$\pm$&\hspace{-2.4em}$0.114$ \\
	\hline
\end{tabular}\\[0.5em]
{\bf Note.} All the uncertainties are $1\sigma$ for the different data sets.\hfill${}$
\end{table*}

\section{Parameter Estimation}
\label{sec:paraest}

In order to find the best-fitting set of parameters for the lensing potential, we shall adopt the maximum likelihood estimator method as a fitting criterion \citep{Hald1999}.  Using the log-likelihood function of the form,
\begin{equation}
	\label{eq:loglikelihood}
	\log\mathcal{L}\left[\tilde{\psi};\boldsymbol{X}\right] = - \frac{1}{2} \sum_{i, j}^{M} \left( \tilde{\psi}(x_i; \boldsymbol{X}) - \tilde{\psi}_i^D\right) \left(\boldsymbol{\Sigma}^{\rm JK}\right)^{-1}_{ij} \left( \tilde{\psi}(x_j; \boldsymbol{X}) - \tilde{\psi}_j^D\right)
\end{equation}
where $\log\mathcal{L}\left[\tilde{\psi}; \boldsymbol{X}\right]$ is the log-likelihood functional of the lensing potential $\tilde{\psi}(x_i; \boldsymbol{X})$ and $\tilde{\psi}_i^D$ is the lensing potential from the data as in Eq.~(\ref{eq:lensingData}).  $\boldsymbol{\Sigma}^{\rm JK}$ is the data covariance matrix in Eq.~(\ref{eq:jackknife}).  The summation is running over all the data points $M$.  The best-fit parameters will be the set of parameters $\boldsymbol{X}_{\text{best}}$ that maximizes the functional.

In order to explore the parameter space effectively, a Markov chain Monte Carlo (MCMC) method with the Metropolis-Hastings algorithm is implemented \citep{Metropolis_ea1953, Hastings1970}.  The Metropolis-Hasting algorithm will advance the state with an acceptance probability from a parameter state $\boldsymbol{X}$ to a parameter state $\boldsymbol{Y}$ given by
\begin{equation}
	\alpha\left(\boldsymbol{X}, \boldsymbol{Y}\right) = \min\left\{1, \frac{\pi(\boldsymbol{Y})\ q(\boldsymbol{Y}, \boldsymbol{X})}{\pi(\boldsymbol{X})\ q(\boldsymbol{X}, \boldsymbol{Y})}\right\},
\end{equation}
where $q(\boldsymbol{Y}, \boldsymbol{X})$ is the proposal probability distribution from $\boldsymbol{X}$ to $\boldsymbol{Y}$.  We shall take a multivariate normal distribution as our density proposal distribution,
\begin{eqnarray}
	\nonumber q(\boldsymbol{Y}, \boldsymbol{X}) &=& \frac{1}{\left(\left(2\pi\right)^N \det\left(\boldsymbol{\Sigma}\right)\right)^{1/2}} \exp\left\{-\frac{1}{2}\left(\boldsymbol{Y} - \boldsymbol{X}\right)^\top \boldsymbol{\Sigma}^{-1} \left(\boldsymbol{Y} - \boldsymbol{X}\right)\right\}, \\
	&\equiv& \mathcal{N}\left(\boldsymbol{Y};\boldsymbol{X}, \boldsymbol{\Sigma}\right),
\end{eqnarray}
where $\boldsymbol{\Sigma}$ is the covariance matrix.  Since our proposal distribution is symmetric, our acceptance probability is $\displaystyle\alpha = \min\left\{1, \pi\left(\boldsymbol{Y}\right)/\pi\left(\boldsymbol{X}\right) \right\}$.  $\pi(\boldsymbol{X})$ is the weighing distribution, which shall be taken as the likelihood function, i.e. the exponential of Eq.~(\ref{eq:loglikelihood}).

The MCMC will sample the parameter space giving a chain, $\left\{\boldsymbol{X}_i, i=0,1,\ldots\right\}$, for the $i$th iteration.  The chain will continue until an equilibrium state is reached, and the MCMC will sample the underlying posterior distribution.  For an MCMC with a fixed variance of the proposal distribution $\boldsymbol{\Sigma}$, this could lead to a situation where the acceptance rate is either too small or too large.  This could result in a final posterior distribution being to localized or a slow convergence rate, respectively.  For a better learning performance, an adaptive MCMC algorithm \citep{Andrieu_Thoms2008} shall be implemented.  The algorithm has the advantage of adjusting the variance according to the the acceptance probability by introducing the scaling factor $\lambda^j$ for the marginal variance $\left[\boldsymbol{\Sigma}\right]_{j,j}$.  In addition, the mean value of the posterior distribution
\begin{equation}
	\boldsymbol{\mu} \equiv \langle \boldsymbol{X}\rangle,
\end{equation}
and the covariance matrix
\begin{equation}
	\boldsymbol{\Sigma} \equiv \big\langle \left( \boldsymbol{X} - \boldsymbol{\mu}\right)^\top \cdot \left( \boldsymbol{X} - \boldsymbol{\mu} \right)\big\rangle,
\end{equation}
are updated for each iteration.  This is achieved by a comparison of $\alpha(\boldsymbol{X}, \boldsymbol{Y})$ with a preferred acceptance rate $\alpha_{*}$.  If $\alpha(\boldsymbol{X}, \boldsymbol{Y}) < \alpha_{*}$ for most transition attempts, then $\boldsymbol{\Sigma}$ should be increased.  If, on the other hand, $\alpha(\boldsymbol{X}, \boldsymbol{Y}) > \alpha_{*}$ for most transition attempts, then $\boldsymbol{\Sigma}$ should be decreased.  However, the acceptance rate will be compared with the preferred acceptance rate component-wise to allow the case where $\boldsymbol{\Sigma}$ should decrease in one direction in the parameter space and increase on in the other direction.  Hence, our proposal distribution will be given by $\mathcal{N}\left(\boldsymbol{Y};\boldsymbol{X}, \boldsymbol{\Lambda}^{1/2} \boldsymbol{\Sigma} \boldsymbol{\Lambda}^{1/2}\right)$, where $\boldsymbol{\Lambda}$ is a scaling matrix,
\begin{equation}
	\boldsymbol{\Lambda} = \text{diag}\left(\lambda^1, \ldots, \lambda^N\right).
\end{equation}
The algorithm is explained in detail in the following, where $\hat{e}_k$ is a unit vector with zeros everywhere except the $k$th component and $\zeta_{i}$ is a nonincreasing function of $i$, the iteration number:

\begin{enumerate}
	\item \normalsize Initialize $\boldsymbol{X}_0$, $\boldsymbol{\mu}_0$, $\boldsymbol{\Sigma}_0$, and $\lambda_0^1, \ldots, \lambda_0^N$ for $i = 0$.
	\item \normalsize Iterate $i + 1$ using the following procedure:
	\begin{enumerate}
		\item For a given $\boldsymbol{\mu}_i, \boldsymbol{\Sigma}_i$ and $\lambda_i^1, \ldots, \lambda_i^N$ sample $\Delta\boldsymbol{X}_i$ from the distribution $\mathcal{N}\left(\boldsymbol{X}_i, \boldsymbol{\Lambda}_i^{1/2} \boldsymbol{\Sigma}_{i} \boldsymbol{\Lambda}_{i}^{1/2}\right)$.
		\item Propose a new state $\boldsymbol{Y}_{i + 1} = \boldsymbol{X}_i + \Delta\boldsymbol{X}_i$ and transverse to the new state with probability $\alpha\left(\boldsymbol{X}_i, \boldsymbol{Y}_{i+1}\right)$; otherwise, $\boldsymbol{X}_{i+1} = \boldsymbol{X}_i$.
		\item Update the scaling factor for $j = 1,\ldots N$,
		{\begin{align}
			\nonumber\log\left(\lambda_{i+1}^j\right) = \log\left(\lambda_{i}^j\right) + \zeta_{i+1} &\left[ \alpha\left(\boldsymbol{X}_i, \boldsymbol{X}_i + \Delta\boldsymbol{X}_i(j) \hat{e}_j  \right)\right.\\
			&\left. - \alpha_{*}\right]
		\end{align}}
		\item Update mean and covariance matrix,
		\begin{equation}
			\boldsymbol{\mu}_{i+1} = \boldsymbol{\mu}_{i} + \zeta_{i+1} \left( \boldsymbol{X}_{i+1} - \boldsymbol{\mu}_i\right),
		\end{equation}
		and
		{\begin{align}
			\nonumber\boldsymbol{\Sigma}_{i+1} =  \boldsymbol{\Sigma}_{i} + \zeta_{i+1} &\left[\left(\boldsymbol{X}_{i+1} - \boldsymbol{\mu}_i\right)^\top \left(\boldsymbol{X}_{i+1} - \boldsymbol{\mu}_i\right)\right. \\
			&-\left. \boldsymbol{\Sigma}_i \right]
		\end{align}}
	\end{enumerate}
	\item Repeat the procedure until an equilibrium state is achieved.
\end{enumerate}

There are no constraints on the functional form of $\zeta_{i}$ as long as it is nonincreasing \citep{Roberts_Rosenthal2007}.  We shall set $\zeta_i$ as
\begin{equation}
	\zeta_i = \frac{1}{10\ i^{1/3}}.
\end{equation}

\section{Results}
\label{sec:results}

We divide our lensing potential data into three data sets; Data-I where we include the lensing potential from the center to $1\, R_V$, and Data-II and Data-III, where we include the lensing potential to $2\, R_V$ and  $3\, R_V$, respectively.  Data-I will help in exploring the interior structure of voids, while Data-II and Data-III will get an overall fit to the void profile.  With the inclusion of an additional parameter from the HSW void profile, we shall refer to the model without the parameter, $\gamma_V$, as Model A and the model with the environment parameter as Model B.  The initial position in the parameter space for each chain will be randomly selected from the parameter range, as shown in Eq.~(\ref{eq:prior}): 
\begin{eqnarray}
	\label{eq:prior}
	\nonumber\alpha \in \left[0.00, 4.50\right],&\quad& \beta \in \left[ 0.00, 24.00\right], \\ \delta_c \in \left[-0.80, -0.20\right], &\quad& R_S/R_V \in \left[ 0.80, 1.60\right], \\
	\nonumber\gamma_V &\in& \left[-0.30, 0.30 \right].
\end{eqnarray}

The range for each parameter is large enough to ensure that our fitting parameters will fall in an agreement with other works \citep{Hamaus_ea2014a, Hamaus_ea2015}.  In the adaptive MCMC described in \S\ref{sec:paraest}, each dataset was run for a total number of 100 chains with 20,000 sampling points per chain.  As for the burn-in period, we abandon the initial 15,000 points and take only the last 5000 points for our parameter estimation.  
For Data-I our fitting parameters are all the void parameters without the environment parameter, i.e. $\gamma_V = 0$.  This is due to the fact that the fit to Data-I with only void parameters is already good without the environment parameter.  Adding the environment parameter would produce an unnecessary overfit to the data.  However, the void parameters with the environment parameter are used to fit the Data-II and Data-III as the merit of an additional parameter to the model overcomes the penalty of adding more parameters (Occam's razor).  The interpretation of the environment parameter will be further discussed in \S\ref{sec:discussion}.

We take the average mean and covariance matrix as follows:
\begin{equation}
	\boldsymbol{\mu}_\text{ave} = \frac{1}{N_c}\sum_{i=1}^{N_c} \boldsymbol{\mu}_i,
\end{equation}
and
\begin{equation}
	\boldsymbol{\Sigma}_\text{ave}^{-1} = \frac{1}{N_c}\sum_{i=1}^{N_c} \boldsymbol{\Sigma}_i^{-1},
\end{equation}
where $N_c$ is the number of chains and $\boldsymbol{\Sigma}_i$ is the final covariance matrix for the $i$th chain.  We shall use the \texttt{Gaumixmod} algorithm \citep{Press_ea2007} to find the best-fit $\boldsymbol{\mu}_i$ and $\boldsymbol{\Sigma}_i$ for each chain.  Our best-fitting parameters for Model A and Model B are shown in \tref{tab:fittingparams}.  The resulting void profile and lensing potential from the best-fit parameters in each dataset are shown in \fref{fig:combine}.  For comparison, the best-fit parameters for Model A and Model B for each dataset are shown in \fref{fig:compare}.



\section{Discussions and Conclusions}
\label{sec:discussion}

{\color{red}
\begin{figure}\centering
	\includegraphics[scale=0.7]{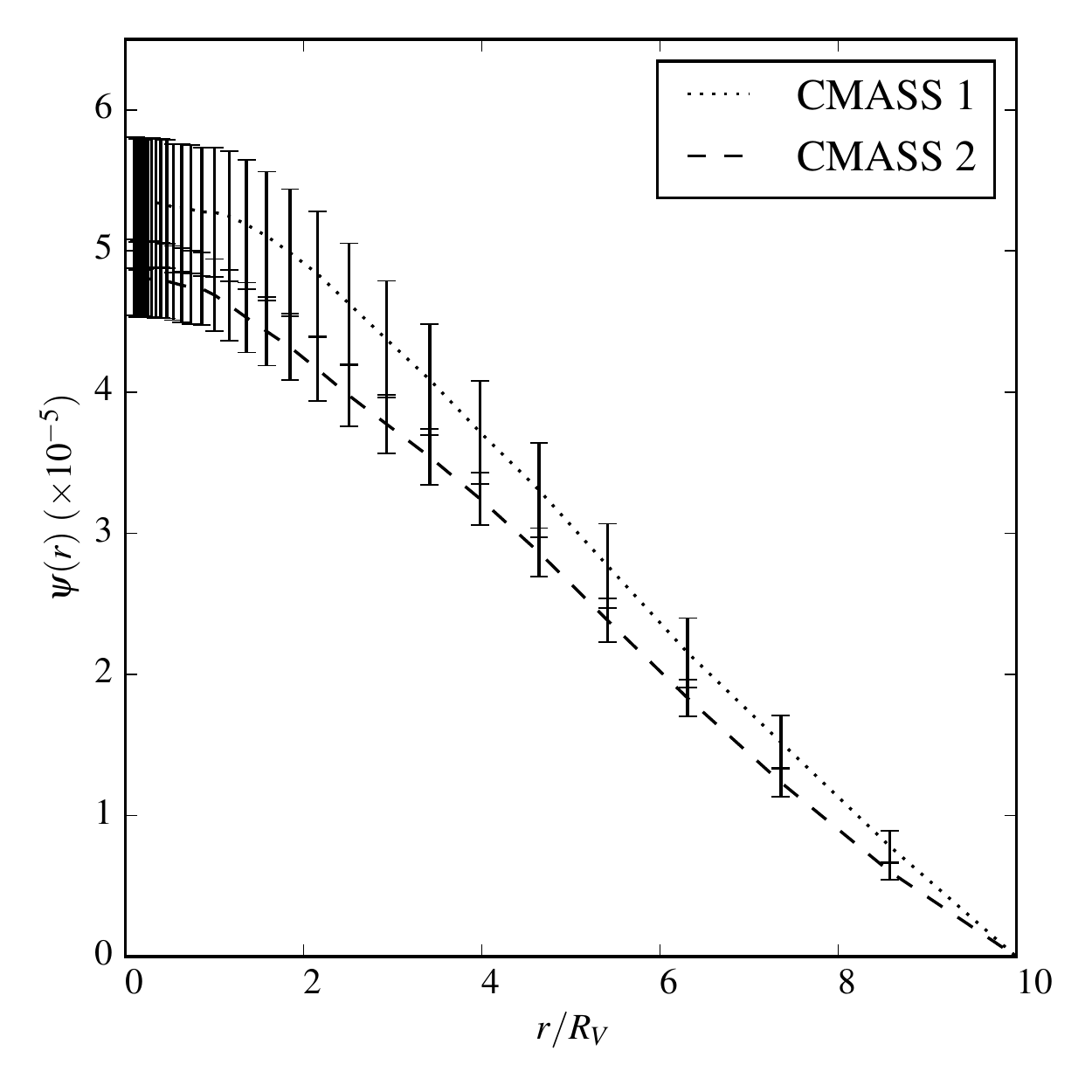}
	\caption{\label{fig:lensp_long} Lensing potential from \textit{Planck} data where voids are detected with SDSS data.}
\end{figure}
}

\begin{figure*}
	\includegraphics[scale=0.7]{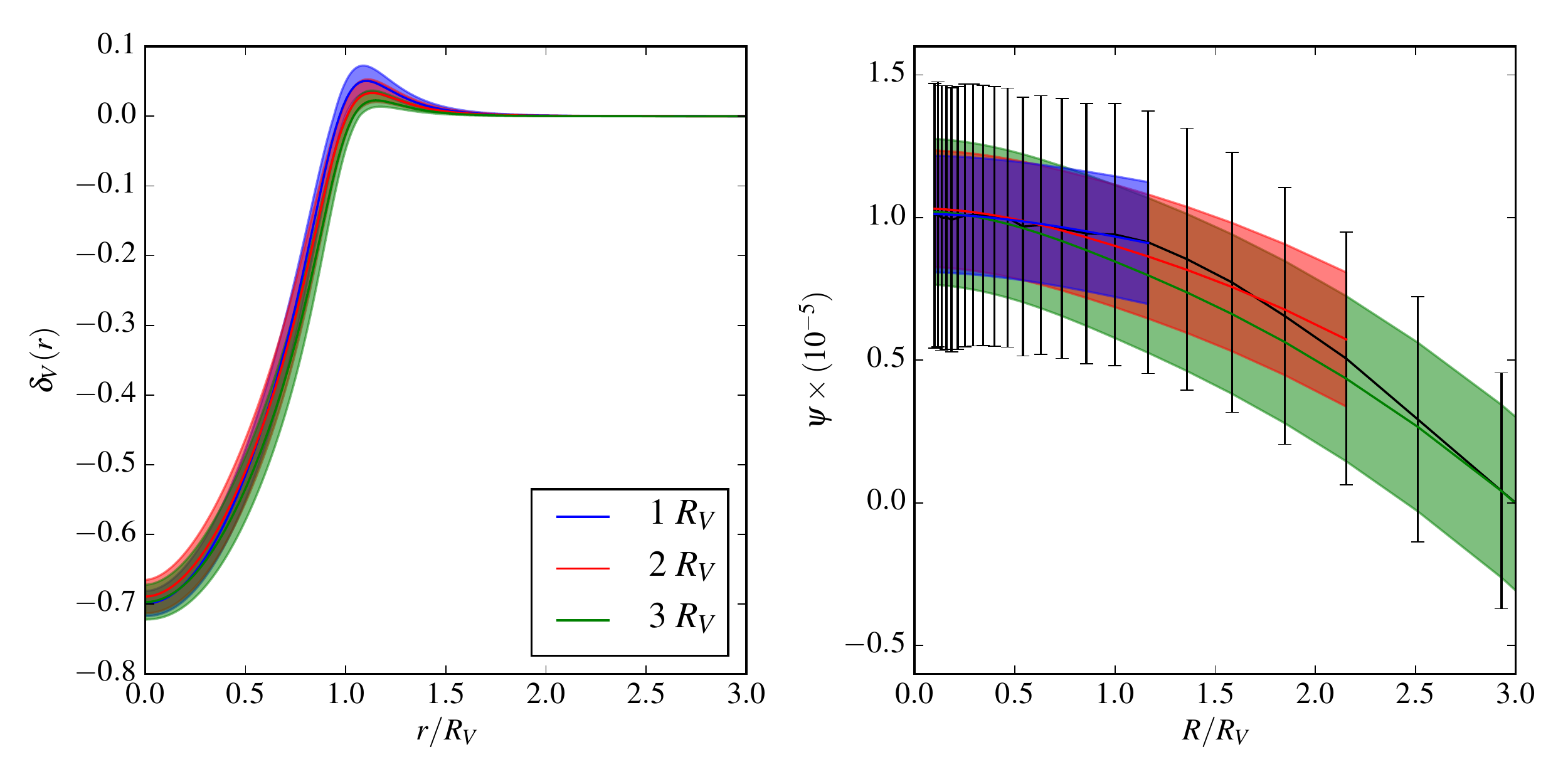}
	\caption{\label{fig:combine} Best-fit void profiles and lensing potentials for each dataset.  The blue, red, and green bands are for Data-I, Data-II, and Data-III respectively.  All the error bars are $1\,\sigma$ for cmass1 sample.}
\end{figure*}

From the void profiles shown in the left panel of \fref{fig:combine} and the void parameters in \tref{tab:fittingparams}, if we calculate the excess mass $\delta m = 4\pi \int_0^\infty \delta_V(x) * x^2 \mbox{d}x$, we could see that the excess mass lies within the range $(-2.5, -1.5)$ for Model A and $(-1.5, -0.5)$ for Model B.  The range of value indicates that voids found in SDSS data are mostly undercompensated, having less density than the average.  This indicates the fact that voids in underdense regions are easily found with high significant levels in SDSS data since large voids are usually found in underdense regions.  We also notice that without $\gamma_V$ the excess mass is higher in the negative value.

%

\begin{figure*}\centering
	\includegraphics[scale=0.8]{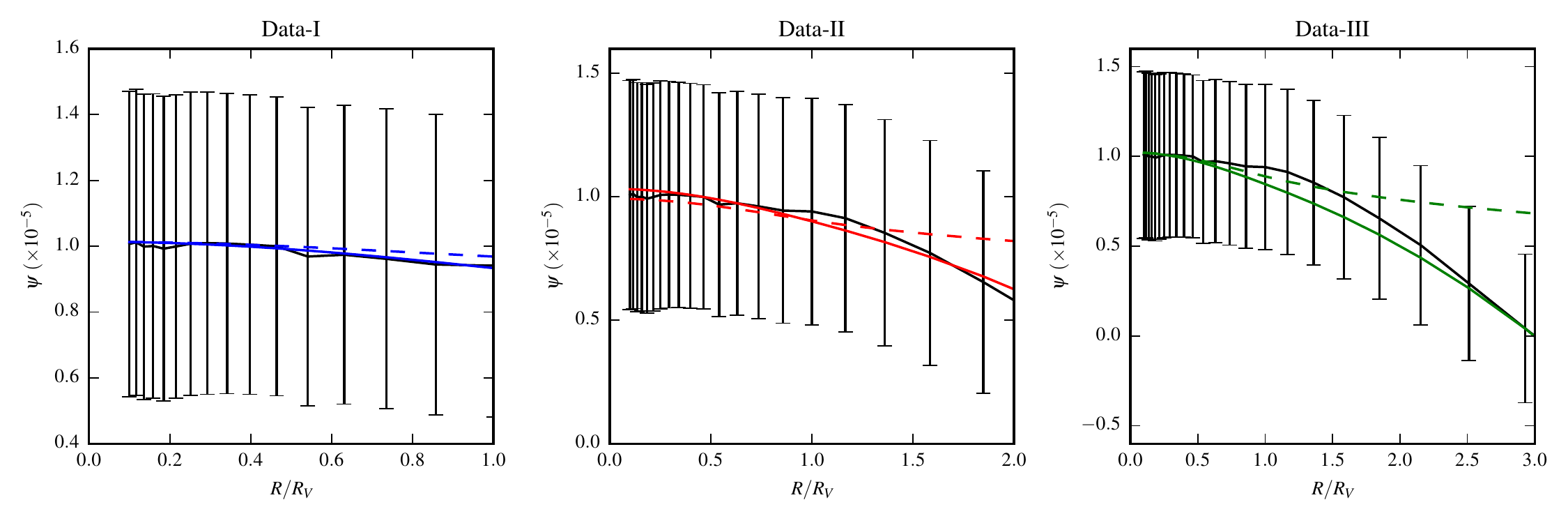}
	\caption{\label{fig:compare} Best-fit lensing potential for Data-I, Data-II, and Data-III (left to right).  The solid lines are void parameter with the environment parameter while the dashed lines are void parameters without the environment parameter.  Even though all the models with the environment parameter fit better with the data, the model without the environment parameter also fits well for Data-I and Data-II.  All the plots are subjected to the condition $\psi(r = 3R_V) = 0$.  The error bars are $1\,\sigma$.}
\end{figure*}

In \tref{tab:fittingparams}, we show the best-fit parameters within $1\,\sigma$ uncertainties.  The parameters from different datasets are all agree within the uncertainty ranges.  The values of the environment parameter $\gamma_V$ are approximately $\sim -0.08$ for cmass1 and $\sim -0.05$ for cmass2 with a high significant detection level from zero.  The negative value of $\gamma_V$ indicates that on average the voids in SDSS data are in a locally underdense region, which is also confirmed by the integrated mass discussed previously.  This effect could be seen from the lensing potential shown in \fref{fig:lensp_long}.  Since the deflection angle is equal to the gradient of the lensing potential $(\boldsymbol{\alpha} \equiv \boldsymbol{\nabla}_\perp \psi)$, the nonzero gradient of the lensing potential at large radius indicates that there is some residue lensing power.  The lensing power mainly comes from the deficit (or excess) mass from the voids (or clusters). The constancy of the gradient translates into the constancy of the excess mass.  Our MCMC has shown that this residue power is caused by the mass deficit by having a negative value of $\gamma_V$.  However, this comes with the caveat, as stated in \cite{PlanckXV_2015}, that the lensing potential has a very red spectrum and when cutting the map into small regions it could cause a leakage issue. This is the reason the team chose to release the lensing convergence 
rather than the potential. And for our purposes we need to convert it back to the lensing potential (\S \ref{ssec:planckdata}).

To justify the necessity of the environment parameter, the best-fit lensing potentials from the models with and without the environment parameter are shown in \fref{fig:compare}.  The likelihood ratios between the models with and without the environment parameter $\mathcal{L}_\text{w}/\mathcal{L}_\text{w/o}$ are 1.01, 1.36, and 77.4 for Data-I, Data-II, and Data-III, respectively, for cmass1.  Without taking the penalty of an additional parameter in $\mathcal{L}_\text{w}$ into account, we could see that the environment parameter is preferred in Data-III, while it is less preferable for Data-I and Data-II.  The best-fit parameters for all samples and models are shown in \tref{tab:fittingparams}.  From Table.~\ref{tab:fittingparams}, the significant detection levels for the environment parameter for Data-I, Data-II and Data-III are approximately $15\sigma$, $30\,\sigma$, and $47\,\sigma$ for Data-I, Data-II, and, Data-III, respectively, for the cmass1 sample respectively.  The general results hold similarly for the cmass2 sample.

To analyze the effect of $\gamma_V$ on the other parameters, we shall take to the best-fit parameters from Data-III for the cmass1 sample from \tref{tab:fittingparams} for a comparison.  The effects of how the HSW parameters alter the shape of the void profile are explicitly shown in Figure.~8 in \cite{Barreira_ea2015}.  In both models, the values of $\beta$ between the two models are not much different given the uncertainties in the values---the values differ by $2.3\,\sigma$.  The similarity in the values of $\beta$ indicates that the extensions of the compensation region are similar.  However, the values of $\alpha$, $\delta_c$, and $R_S/R_V$ are significantly different by $4.4\,\sigma$, $3.6\,\sigma$, and $6.6\,\sigma$, respectively.  $\alpha$ describes the slope of the underdense region, $\delta_c$ describes the depth of the void profile, and $R_S/R_V$ is the zero-crossing radius.  In general, Model B (with $\gamma_V$) gives a shallower void profile than the Model A (without $\gamma_V$) by having smaller values of $\alpha$ and $\delta_c$.  This indicates that $\gamma_V$ has a direct degenerate effect with both $\alpha$ and $\delta_c$---lowering the mean density could be compensated by having a shallower profile.

Attempts to recover or fit the HSW void parameters are found in \citet[hereafter, \hyperlink{ref:n14}{N14}]{Nadathur_ea2014} and \citet[hereafter, \hyperlink{ref:h15}{H15}]{Hamaus_ea2015}.  In \hyperlink{ref:n14}{N14}, the stacked void profiles for different radii and redshifts are compared between a mock luminous red galaxy (LRG) catalog from the Jubilee simulation and the SDSS LRG and Main Galaxy samples.  They have found that the void profiles from the simulations and the SDSS galaxy samples matched.  \hyperlink{ref:h15}{H15} investigated the redshift-space distortions between pairs of galaxies from a mock galaxies catalog with HOD parameters from \cite{Zheng_ea2007} and \cite{Manera_ea2013} and stacked voids in redshift space.  The inference on the HSW void parameters was made by assuming the Gaussian streaming model, where the distribution of the pairwise line-of-sight velocities is assumed to be Gaussian.  The HSW parameters in \hyperlink{ref:n14}{N14} and \hyperlink{ref:h15}{H15} are $(\alpha, \beta, \delta_c, R_S/R_V) = (1.57, 5.72, -0.69, 0.81)$ and $(\alpha, \beta, \delta_c) = (0.96~\pm~0.14, 8.84~\pm~1.16, -0.912~\pm~0.052)$ respectively.  The main differences in the parameter constraints from our work and theirs are from the different methodology used in deriving the parameters and the inclusion of $\gamma_V$.  We also notice that the value of $\delta_c$ between \hyperlink{ref:n14}{N14} (observationally derived) and our value, especially for Data-I, are similar, while the value of $\delta_c$ from \hyperlink{ref:h15}{H15} (mock catalog derived) is remarkably different.  This may indicate a systematic bias between the SDSS samples and the mock catalog.

To summarize this work, we cross-correlated the \textit{Planck} lensing map with 516 voids found in the SDSS data and stacked them to obtain the stacked lensing potential from voids.  From the stacked void lensing potential, we recover the HSW void parameter from three different data sets: Data-I to $1 R_V$, Data-II to $2 R_V$, and Data-III to $3 R_V$.  We have found that it is necessary to include the environment parameter in the void profile to obtain a good fit to the data.  The environment parameter has a physical interpretation that voids found in the SDSS data are mostly undercompensated voids and reside within an underdense region.  The effects of the deficit mass are shown in \fref{fig:lensp_long}, where the gradient of the lensing potential is constant at large distances.  

\acknowledgments

We would like to thank Paul Sutter for his useful comments.  T.C. acknowledges the support from the National Astronomical Research Institute of Thailand (NARIT).  This work is supported by a NARIT research grant and its High Performance Computer Facility.  This work has been done within the Labex ILP (reference ANR-10-LABX-63) part of the Idex SUPER, and received financial support managed by the Agence Nationale de la Recherche, as part of the programme Investissements d’avenir under the reference ANR-11-IDEX- 0004-02.


\begin{thebibliography}{99}
\expandafter\ifx\csname natexlab\endcsname\relax\def\natexlab#1{#1}\fi

\bibitem[{{Abazajian} {et~al.}(2009){Abazajian}, {Adelman-McCarthy},
  {Ag{\"u}eros}, {Allam}, {Allende Prieto}, {An}, {Anderson}, {Anderson},
  {Annis}, {Bahcall}, \& et~al.}]{SDSSDR7}
{Abazajian}, K.~N., {Adelman-McCarthy}, J.~K., {Ag{\"u}eros}, M.~A., {et~al.}
  2009, \apjs, 182, 543

\bibitem[{{Ahn} {et~al.}(2014){Ahn}, {Alexandroff}, {Allende Prieto}, {Anders},
  {Anderson}, {Anderton}, {Andrews}, {Aubourg}, {Bailey}, {Bastien}, \&
  et~al.}]{SDSSDR10}
{Ahn}, C.~P., {Alexandroff}, R., {Allende Prieto}, C., {et~al.} 2014, \apjs,
  211, 17

\bibitem[{{Alcock} \& {Paczynski}(1979)}]{Alcock_Paczynski1979}
{Alcock}, C., \& {Paczynski}, B. 1979, \nat, 281, 358

\bibitem[{Andrieu \& Thoms(2008)}]{Andrieu_Thoms2008}
Andrieu, C., \& Thoms, J. 2008, Statistics and Computing, 18, 343

\bibitem[{{Barreira} {et~al.}(2015){Barreira}, {Cautun}, {Li}, {Baugh}, \&
  {Pascoli}}]{Barreira_ea2015}
{Barreira}, A., {Cautun}, M., {Li}, B., {Baugh}, C.~M., \& {Pascoli}, S. 2015,
  \jcap, 8, 028

\bibitem[{{Bartelmann} \& {Schneider}(2001)}]{Bartelmann_Schneider2001}
{Bartelmann}, M., \& {Schneider}, P. 2001, \physrep, 340, 291

\bibitem[{{Beygu} {et~al.}(2013){Beygu}, {Kreckel}, {van de Weygaert}, {van der
  Hulst}, \& {van Gorkom}}]{Beygu_ea2013}
{Beygu}, B., {Kreckel}, K., {van de Weygaert}, R., {van der Hulst}, J.~M., \&
  {van Gorkom}, J.~H. 2013, \aj, 145, 120

\bibitem[{{Biswas} {et~al.}(2010){Biswas}, {Alizadeh}, \&
  {Wandelt}}]{Biswas_ea2010}
{Biswas}, R., {Alizadeh}, E., \& {Wandelt}, B.~D. 2010, \prd, 82, 023002

\bibitem[{{Bolejko} {et~al.}(2013){Bolejko}, {Clarkson}, {Maartens}, {Bacon},
  {Meures}, \& {Beynon}}]{Bolejko_ea2013}
{Bolejko}, K., {Clarkson}, C., {Maartens}, R., {et~al.} 2013, Physical Review
  Letters, 110, 021302

\bibitem[{{Bos} {et~al.}(2012){Bos}, {van de Weygaert}, {Dolag}, \&
  {Pettorino}}]{Bos_ea2012}
{Bos}, E.~G.~P., {van de Weygaert}, R., {Dolag}, K., \& {Pettorino}, V. 2012,
  \mnras, 426, 440

\bibitem[{{Boylan-Kolchin} {et~al.}(2009){Boylan-Kolchin}, {Springel}, {White},
  {Jenkins}, \& {Lemson}}]{Boylan-Kolchin_ea2009}
{Boylan-Kolchin}, M., {Springel}, V., {White}, S.~D.~M., {Jenkins}, A., \&
  {Lemson}, G. 2009, \mnras, 398, 1150

\bibitem[{{Cai} {et~al.}(2014){Cai}, {Neyrinck}, {Szapudi}, {Cole}, \&
  {Frenk}}]{Cai_ea2014}
{Cai}, Y.-C., {Neyrinck}, M.~C., {Szapudi}, I., {Cole}, S., \& {Frenk}, C.~S.
  2014, \apj, 786, 110

\bibitem[{{Cai} {et~al.}(2015){Cai}, {Padilla}, \& {Li}}]{Cai_ea2015}
{Cai}, Y.-C., {Padilla}, N., \& {Li}, B. 2015, \mnras, 451, 1036

\bibitem[{{Ceccarelli} {et~al.}(2006){Ceccarelli}, {Padilla}, {Valotto}, \&
  {Lambas}}]{Ceccarelli_ea2006}
{Ceccarelli}, L., {Padilla}, N.~D., {Valotto}, C., \& {Lambas}, D.~G. 2006,
  \mnras, 373, 1440

\bibitem[{{Chantavat} {et~al.}(2016){Chantavat}, {Sawangwit}, {Sutter}, \&
  {Wandelt}}]{Chantavat_ea2016}
{Chantavat}, T., {Sawangwit}, U., {Sutter}, P.~M., \& {Wandelt}, B.~D. 2016,
  \prd, 93, 043523

\bibitem[{{Chen} \& {Kantowski}(2015)}]{Chen_Kantowski2015}
{Chen}, B., \& {Kantowski}, R. 2015, \prd, 91, 083014

\bibitem[{{Chen} {et~al.}(2015){Chen}, {Kantowski}, \& {Dai}}]{Chen_ea2015}
{Chen}, B., {Kantowski}, R., \& {Dai}, X. 2015, \apj, 804, 130

\bibitem[{{Clampitt} {et~al.}(2013){Clampitt}, {Cai}, \&
  {Li}}]{Clampitt_ea2013}
{Clampitt}, J., {Cai}, Y.-C., \& {Li}, B. 2013, \mnras, 431, 749

\bibitem[{{Clampitt} \& {Jain}(2015)}]{Clampitt_Jain2015}
{Clampitt}, J., \& {Jain}, B. 2015, \mnras, 454, 3357

\bibitem[{{Courtois} {et~al.}(2012){Courtois}, {Hoffman}, {Tully}, \&
  {Gottl{\"o}ber}}]{Courtois_ea2012}
{Courtois}, H.~M., {Hoffman}, Y., {Tully}, R.~B., \& {Gottl{\"o}ber}, S. 2012,
  \apj, 744, 43

\bibitem[{{Das} \& {Spergel}(2009)}]{Das_Spergel2009}
{Das}, S., \& {Spergel}, D.~N. 2009, \prd, 79, 043007

\bibitem[{{Einstein}(1936)}]{Einstein_1936}
{Einstein}, A. 1936, Science, 84, 506

\bibitem[{{G{\'o}rski} {et~al.}(2005){G{\'o}rski}, , \& et~al.}]{HEALPixref}
{G{\'o}rski}, K.~M., , \& et~al. 2005, \apj, 622, 759

\bibitem[{{Granett} {et~al.}(2008){Granett}, {Neyrinck}, \&
  {Szapudi}}]{Granett_ea2008}
{Granett}, B.~R., {Neyrinck}, M.~C., \& {Szapudi}, I. 2008, \apjl, 683, L99

\bibitem[{{Gruen} {et~al.}(2016){Gruen}, {Friedrich}, {Amara}, {Bacon},
  {Bonnett}, {Hartley}, {Jain}, {Jarvis}, {Kacprzak}, {Krause}, {Mana}, {Rozo},
  {Rykoff}, {Seitz}, {Sheldon}, {Troxel}, {Vikram}, {Abbott}, {Abdalla},
  {Allam}, {Armstrong}, {Banerji}, {Bauer}, {Becker}, {Benoit-L{\'e}vy},
  {Bernstein}, {Bernstein}, {Bertin}, {Bridle}, {Brooks}, {Buckley-Geer},
  {Burke}, {Capozzi}, {Carnero Rosell}, {Carrasco Kind}, {Carretero}, {Crocce},
  {Cunha}, {D'Andrea}, {da Costa}, {DePoy}, {Desai}, {Diehl}, {Dietrich},
  {Doel}, {Eifler}, {Neto}, {Fernandez}, {Flaugher}, {Fosalba}, {Frieman},
  {Gerdes}, {Gruendl}, {Gutierrez}, {Honscheid}, {James}, {Kuehn},
  {Kuropatkin}, {Lahav}, {Li}, {Lima}, {Maia}, {March}, {Martini}, {Melchior},
  {Miller}, {Miquel}, {Mohr}, {Nord}, {Ogando}, {Plazas}, {Reil}, {Romer},
  {Roodman}, {Sako}, {Sanchez}, {Scarpine}, {Schubnell}, {Sevilla-Noarbe},
  {Smith}, {Soares-Santos}, {Sobreira}, {Suchyta}, {Swanson}, {Tarle},
  {Thaler}, {Thomas}, {Walker}, {Wechsler}, {Weller}, {Zhang}, \&
  {Zuntz}}]{Gruen_ea2016}
{Gruen}, D., {Friedrich}, O., {Amara}, A., {et~al.} 2016, \mnras, 455, 3367

\bibitem[{{Hald}(1999)}]{Hald1999}
{Hald}, A. 1999, Statist. Sci., 14, 214

\hypertarget{ref:h15}{
\bibitem[{{Hamaus} {et~al.}(2015){Hamaus}, {Sutter}, {Lavaux}, \&
  {Wandelt}}]{Hamaus_ea2015}
{Hamaus}, N., {Sutter}, P.~M., {Lavaux}, G., \& {Wandelt}, B.~D. 2015, \jcap,
  11, 036}

\bibitem[{{Hamaus} {et~al.}(2014{\natexlab{a}}){Hamaus}, {Sutter}, \&
  {Wandelt}}]{Hamaus_ea2014a}
{Hamaus}, N., {Sutter}, P.~M., \& {Wandelt}, B.~D. 2014{\natexlab{a}}, Physical
  Review Letters, 112, 251302

\bibitem[{{Hamaus} {et~al.}(2014{\natexlab{b}}){Hamaus}, {Wandelt}, {Sutter},
  {Lavaux}, \& {Warren}}]{Hamaus_ea2014b}
{Hamaus}, N., {Wandelt}, B.~D., {Sutter}, P.~M., {Lavaux}, G., \& {Warren},
  M.~S. 2014{\natexlab{b}}, Physical Review Letters, 112, 041304

\bibitem[{{Hastings}(1970)}]{Hastings1970}
{Hastings}, W.~K. 1970, Biometrika, 57, 97

\bibitem[{{Higuchi} {et~al.}(2013){Higuchi}, {Oguri}, \&
  {Hamana}}]{Higuchi_ea2013}
{Higuchi}, Y., {Oguri}, M., \& {Hamana}, T. 2013, \mnras, 432, 1021

\bibitem[{{Hinshaw} {et~al.}(2013){Hinshaw}, {Larson}, {Komatsu}, {Spergel},
  {Bennett}, {Dunkley}, {Nolta}, {Halpern}, {Hill}, {Odegard}, {Page}, {Smith},
  {Weiland}, {Gold}, {Jarosik}, {Kogut}, {Limon}, {Meyer}, {Tucker}, {Wollack},
  \& {Wright}}]{Hinshaw_ea2013}
{Hinshaw}, G., {Larson}, D., {Komatsu}, E., {et~al.} 2013, \apjs, 208, 19

\bibitem[{{Hotchkiss} {et~al.}(2015){Hotchkiss}, {Nadathur}, {Gottl{\"o}ber},
  {Iliev}, {Knebe}, {Watson}, \& {Yepes}}]{Hotchkiss_ea2015}
{Hotchkiss}, S., {Nadathur}, S., {Gottl{\"o}ber}, S., {et~al.} 2015, \mnras,
  446, 1321

\bibitem[{{Ili{\'c}} {et~al.}(2013){Ili{\'c}}, {Langer}, \&
  {Douspis}}]{Ilic_ea2013}
{Ili{\'c}}, S., {Langer}, M., \& {Douspis}, M. 2013, \aap, 556, A51

\bibitem[{{Jennings} {et~al.}(2013){Jennings}, {Li}, \& {Hu}}]{Jennings_ea2013}
{Jennings}, E., {Li}, Y., \& {Hu}, W. 2013, \mnras, 434, 2167

\bibitem[{{Kaiser}(1987)}]{Kaiser_1987}
{Kaiser}, N. 1987, \mnras, 227, 1

\bibitem[{{Krause} {et~al.}(2013){Krause}, {Chang}, {Dor{\'e}}, \&
  {Umetsu}}]{Krause_ea2013}
{Krause}, E., {Chang}, T.-C., {Dor{\'e}}, O., \& {Umetsu}, K. 2013, \apjl, 762,
  L20

\bibitem[{{Lavaux} \& {Wandelt}(2012)}]{Lavaux_Wandelt2012}
{Lavaux}, G., \& {Wandelt}, B.~D. 2012, \apj, 754, 109

\bibitem[{{Lee} \& {Park}(2009)}]{Lee_Park2009}
{Lee}, J., \& {Park}, D. 2009, \apjl, 696, L10

\bibitem[{{Manera} {et~al.}(2013){Manera}, {Scoccimarro}, {Percival},
  {Samushia}, {McBride}, {Ross}, {Sheth}, {White}, {Reid}, {S{\'a}nchez}, {de
  Putter}, {Xu}, {Berlind}, {Brinkmann}, {Maraston}, {Nichol}, {Montesano},
  {Padmanabhan}, {Skibba}, {Tojeiro}, \& {Weaver}}]{Manera_ea2013}
{Manera}, M., {Scoccimarro}, R., {Percival}, W.~J., {et~al.} 2013, \mnras, 428,
  1036

\bibitem[{{Melchior} {et~al.}(2014){Melchior}, {Sutter}, {Sheldon}, {Krause},
  \& {Wandelt}}]{Melchior_ea2014}
{Melchior}, P., {Sutter}, P.~M., {Sheldon}, E.~S., {Krause}, E., \& {Wandelt},
  B.~D. 2014, \mnras, 440, 2922

\bibitem[{{Metropolis} {et~al.}(1953){Metropolis}, {Rosenbluth}, {Rosenbluth},
  {Teller}, \& {Teller}}]{Metropolis_ea1953}
{Metropolis}, N., {Rosenbluth}, A.~W., {Rosenbluth}, M.~N., {Teller}, A.~H., \&
  {Teller}, E. 1953, \jcp, 21, 1087

\hypertarget{ref:n14}{
\bibitem[{{Nadathur}(2016)}]{Nadathur_2016}
{Nadathur}, S. 2016, \mnras, 461, 358}

\bibitem[{{Nadathur} {et~al.}(2016){Nadathur}, {Hotchkiss}, {Diego}, {Iliev},
  {Gottl{\"o}ber}, {Watson}, \& {Yepes}}]{Nadathur_ea2014}
{Nadathur}, S., {Hotchkiss}, S., {Diego}, J.~M., {et~al.} 2016, in IAU
  Symposium, Vol. 308, The Zeldovich Universe: Genesis and Growth of the Cosmic
  Web, ed. R.~{van de Weygaert}, S.~{Shandarin}, E.~{Saar}, \& J.~{Einasto},
  542--545

\bibitem[{{Neyrinck}(2008)}]{Neyrinck_2008}
{Neyrinck}, M.~C. 2008, \mnras, 386, 2101

\bibitem[{{Okamoto} \& {Hu}(2003)}]{Okamoto_Hu2003}
{Okamoto}, T., \& {Hu}, W. 2003, \prd, 67, 083002

\bibitem[{{Pan} {et~al.}(2012){Pan}, {Vogeley}, {Hoyle}, {Choi}, \&
  {Park}}]{Pan_ea2012}
{Pan}, D.~C., {Vogeley}, M.~S., {Hoyle}, F., {Choi}, Y.-Y., \& {Park}, C. 2012,
  \mnras, 421, 926

\bibitem[{{Penny} {et~al.}(2015){Penny}, {Brown}, {Pimbblet}, {Cluver},
  {Croton}, {Owers}, {Lange}, {Alpaslan}, {Baldry}, {Bland-Hawthorn}, {Brough},
  {Driver}, {Holwerda}, {Hopkins}, {Jarrett}, {Jones}, {Kelvin},
  {Lara-L{\'o}pez}, {Liske}, {L{\'o}pez-S{\'a}nchez}, {Loveday}, {Meyer},
  {Norberg}, {Robotham}, \& {Rodrigues}}]{Penny_ea2015}
{Penny}, S.~J., {Brown}, M.~J.~I., {Pimbblet}, K.~A., {et~al.} 2015, \mnras,
  453, 3519

\bibitem[{{Planck Collaboration} {et~al.}(2014){Planck Collaboration}, {Ade},
  {Aghanim}, {Armitage-Caplan}, {Arnaud}, {Ashdown}, {Atrio-Barandela},
  {Aumont}, {Baccigalupi}, {Banday}, \& et~al.}]{PlanckXVI}
{Planck Collaboration}, {Ade}, P.~A.~R., {Aghanim}, N., {et~al.} 2014, \aap,
  571, A16

\bibitem[{{Planck Collaboration} {et~al.}(2016{\natexlab{a}}){Planck
  Collaboration}, {Adam}, {Ade}, {Aghanim}, {Akrami}, {Alves}, {Arg{\"u}eso},
  {Arnaud}, {Arroja}, {Ashdown}, \& et~al.}]{PlanckI_2015}
{Planck Collaboration}, {Adam}, R., {Ade}, P.~A.~R., {et~al.}
  2016{\natexlab{a}}, \aap, 594, A1

\bibitem[{{Planck Collaboration} {et~al.}(2016{\natexlab{b}}){Planck
  Collaboration}, {Adam}, {Ade}, {Aghanim}, {Arnaud}, {Ashdown}, {Aumont},
  {Baccigalupi}, {Banday}, {Barreiro}, \& et~al.}]{PlanckIX_2015}
---. 2016{\natexlab{b}}, \aap, 594, A9

\bibitem[{{Planck Collaboration} {et~al.}(2016{\natexlab{c}}){Planck
  Collaboration}, {Ade}, {Aghanim}, {Arnaud}, {Ashdown}, {Aumont},
  {Baccigalupi}, {Banday}, {Barreiro}, {Bartlett}, \& et~al.}]{PlanckXIII_2015}
{Planck Collaboration}, {Ade}, P.~A.~R., {Aghanim}, N., {et~al.}
  2016{\natexlab{c}}, \aap, 594, A13

\bibitem[{{Planck Collaboration} {et~al.}(2016{\natexlab{d}}){Planck
  Collaboration}, {Ade}, {Aghanim}, {Arnaud}, {Ashdown}, {Aumont},
  {Baccigalupi}, {Banday}, {Barreiro}, {Bartlett}, \& et~al.}]{PlanckXV_2015}
---. 2016{\natexlab{d}}, \aap, 594, A15

\bibitem[{{Planck Collaboration} {et~al.}(2016{\natexlab{e}}){Planck
  Collaboration}, {Ade}, {Aghanim}, {Arnaud}, {Ashdown}, {Aumont},
  {Baccigalupi}, {Banday}, {Barreiro}, {Bartolo}, \& et~al.}]{PlanckXXI_2015}
---. 2016{\natexlab{e}}, \aap, 594, A21

\bibitem[{{Platen} {et~al.}(2007){Platen}, {van de Weygaert}, \&
  {Jones}}]{Platen_ea2007}
{Platen}, E., {van de Weygaert}, R., \& {Jones}, B.~J.~T. 2007, \mnras, 380,
  551

\bibitem[{Press {et~al.}(2007)Press, Teukolsky, Vetterling, \&
  Flannery}]{Press_ea2007}
Press, W.~H., Teukolsky, S.~A., Vetterling, W.~T., \& Flannery, B.~P. 2007,
  Numerical Recipes: The Art of Scientific Computing, 3rd edn. (New York, NY,
  USA: Cambridge University Press)

\bibitem[{{Roberts} \& {Rosenthal}(2007)}]{Roberts_Rosenthal2007}
{Roberts}, G.~O., \& {Rosenthal}, J.~S. 2007, Journal of Applied Probability,
  44, 458

\bibitem[{{Sachs} \& {Wolfe}(1967)}]{Sachs_Wolfe1967}
{Sachs}, R.~K., \& {Wolfe}, A.~M. 1967, \apj, 147, 73

\bibitem[{{Sheth} \& {van de Weygaert}(2004)}]{Sheth_vandeWeygaert2004}
{Sheth}, R.~K., \& {van de Weygaert}, R. 2004, \mnras, 350, 517

\bibitem[{{Sutter} {et~al.}(2012{\natexlab{a}}){Sutter}, {Lavaux}, {Wandelt},
  \& {Weinberg}}]{Sutter_ea2012b}
{Sutter}, P.~M., {Lavaux}, G., {Wandelt}, B.~D., \& {Weinberg}, D.~H.
  2012{\natexlab{a}}, \apj, 761, 187

\bibitem[{{Sutter} {et~al.}(2012{\natexlab{b}}){Sutter}, {Lavaux}, {Wandelt},
  \& {Weinberg}}]{Sutter_ea2012a}
---. 2012{\natexlab{b}}, \apj, 761, 44

\bibitem[{{Sutter} {et~al.}(2014{\natexlab{a}}){Sutter}, {Lavaux}, {Wandelt},
  {Weinberg}, {Warren}, \& {Pisani}}]{Sutter_ea2014a}
{Sutter}, P.~M., {Lavaux}, G., {Wandelt}, B.~D., {et~al.} 2014{\natexlab{a}},
  \mnras, 442, 3127

\bibitem[{{Sutter} {et~al.}(2014{\natexlab{b}}){Sutter}, {Pisani}, {Wandelt},
  \& {Weinberg}}]{Sutter_ea2014b}
{Sutter}, P.~M., {Pisani}, A., {Wandelt}, B.~D., \& {Weinberg}, D.~H.
  2014{\natexlab{b}}, \mnras, 443, 2983

\bibitem[{{Zheng} {et~al.}(2007){Zheng}, {Coil}, \& {Zehavi}}]{Zheng_ea2007}
{Zheng}, Z., {Coil}, A.~L., \& {Zehavi}, I. 2007, \apj, 667, 760

\bibitem[{{Zivick} {et~al.}(2015){Zivick}, {Sutter}, {Wandelt}, {Li}, \&
  {Lam}}]{Zivick_ea2015}
{Zivick}, P., {Sutter}, P.~M., {Wandelt}, B.~D., {Li}, B., \& {Lam}, T.~Y.
  2015, \mnras, 451, 4215

\end{thebibliography}

\end{document}